\documentclass[twocolumn]{pasj00}

\begin{document}
\SetRunningHead{Tanaka et al.}{Spectral Evolution of the Unusual Slow Nova V5558~Sgr}
\Received{2011/03/16}
\Accepted{2011/05/07}

\title{Spectral Evolution of the Unusual Slow Nova V5558~Sgr}

\author{Jumpei \textsc{Tanaka}$^1$, Daisaku \textsc{Nogami}$^2$, Mitsugu \textsc{Fujii}$^3$, Kazuya \textsc{Ayani}$^4$, Taichi \textsc{Kato}$^1$, Hiroyuki \textsc{Maehara}$^2$, Seiichiro \textsc{Kiyota}$^5$ and Kazuhiro \textsc{Nakajima}$^6$}
\affil{$^1$Dept. of Astronomy, Kyoto University, Sakyo-ku, Kyoto 606-8502}
\affil{$^2$Kwasan Observatory, Kyoto University, Yamashina-ku, Kyoto 607-8471}
\affil{$^3$Fujii-Bisei Observatory, 4500 Kurosaki, Tamashima, Kurashiki, Okayama 713-8126}
\affil{$^4$Bisei Astronomical Observatory, 1723-70 Okura, Bisei-cho, Ibara, Okayama 714-1411}
\affil{$^5$Variable Star Observers League in Japan (VSOLJ), 405-1003 Matsushiro, Tsukuba, Ibaraki 305-0035}
\affil{$^6$Variable Star Observers League in Japan (VSOLJ), 124 Teradani Isato-cho, Kumano, Mie 519-4673}

\KeyWords{stars: individual (V5558 Sagittarii) --- novae, cataclysmic variables} 

\maketitle

\begin{abstract}
We report on the spectral evolution of the enigmatic, very slow nova V5558~Sgr, based on the low-resolution spectra obtained at the Fujii-Bisei Observatory and the Bisei Astronomical Observatory, Japan during a period of 2007 April 6 to 2008 May 3. V5558~Sgr shows a pre-maximum halt and then several flare-like rebrightenings, which is similar to another very slow nova V723~Cas. In our observations, the spectral type of V5558~Sgr evolved from the He/N type toward the Fe~II type during the pre-maximum halt, and then toward the He/N type again. This course of spectral transition was observed for the first time in the long history of the nova research. In the rebrightening stage after the initial brightness maximum, we could identify many emission lines accompanied by a stronger absorption component of the P-Cygni profile at the brightness maxima. We found that the velocity of the P-Cygni absorption component measured from the emission peak decreased at the brightness maxima. Furthermore, we compared the spectra of V5558 Sgr with V723 Cas, and other novae which exhibited several rebrightenings during the early phase.

\end{abstract}

\section{Introduction}
Novae are a kind of cataclysmic variable stars, i.e. close binary systems consisting of a white dwarf and a low-mass normal star (for a review, see \cite{Hellier2001}). The cause of the nova eruption is considered to be the thermonuclear runaway reaction on the surface of the white dwarf. Matters transferred from the secondary are accumulated and compressed on the white dwarf. When the temperature and density increase sufficiently, thermo-nuclear runaway reactions occur to cause the nova eruption (\cite{Prialnik1986}).

The nova eruption is a very exciting phenomenon, and the nova systems are the best classically known transient objects.
Payne-Gaposchkin (1964) classified the nova light curve based on the time in days for the decline by 2 magnitudes from the maximum, $t_2$. 
Duerbeck (1981) reported that there are some light curve types of novae. They suggested a classification system for nova light curves including the class which shows a dust dip. 
Besides these classifications based on the light curve, McLaughlin (1942, 1944) introduced the various classes of the spectral evolution (see also Payne-Gaposchkin~(1957)).
Williams (1992) divided the early post-outburst spectra of novae into two classes, the Fe~II type nova and He/N type nova, based on the emission lines. They proposed that the Fe~II type spectrum is formed in a continuous wind, while the He/N type spectrum is formed in a discrete shell. They also suggested the physics of the hybrid nova, showing the transition from the Fe~II to the He/N spectrum, or simultaneous emission lines of both types.
The diversity of light curves and spectral evolutions of the novae (demonstrated by e.g. \cite{Kiyota2004} and \cite{Strope2010}), however, has not been fully understood. Recently, Strope et al.~(2010) reported the classification and properties of the nova light curves. They noted that there are at least 14 novae showing the rebrightenings during the early phase as V5558 Sgr, and that some of these novae are not very slow novae, in 93 novae of their sample.

V5558~Sgr was discovered by Y. Sakurai at 2007 April 14.777 (UT) at magnitude 10.3. Sakurai gave the position for the new object as R.A. = \timeform{18h10m18s.4}, Decl. = \timeform{-18D46'51"} (see \cite{Nakano2007}). In addition, Sakurai reported that nothing is visible at this location at April 9.8 (limiting magnitude 11.4). Haseda located several prediscovery images of V5558 Sgr, providing the following estimated magnitudes: 11.8, 11.2, 10.8 and 10.4 mag at April 7.780 (UT), 11.792, 13.793 and 14.777, respectively (data taken from CBET 931) (\cite{Nakano2007}).
 Iijima (2007) obtained medium-dispersion spectra at April 20.12 and 20.14 (UT). In their observations, there are Balmer, He~I, Fe~II, Mg~II, Si~II and probably N~II lines in emission. The full widths at half maximum (FWHMs) of H$\beta$ and He~I are 480 km s$^{-1}$ and 360 km s$^{-1}$, respectively. Moreover, most of the prominent emission lines are accompanied by weak absorption components of the P-Cygni profiles blueshifted by 400-500 km s$^{-1}$.

Poggiani (2008a) reported the spectroscopic monitoring of V5558 Sgr during the pre-maximum stage and the early decline, and found that there are Balmer and Fe~II lines in emission during the pre-maximum stage. The spectra after the first maximum showed Balmer, Fe~II and He~I emission lines. Poggiani~(2008a) pointed out that the spectroscopic evolution is similar to that of the slow nova V723~Cas.

Furthermore, Poggiani~(2010) reported the spectroscopic follow-up during the decline stage. They derived the absolute magnitude of $-$6.3 to $-$5.9, the typical magnitude at maximum for slow novae such as V723~Cas and HR~Del. They also estimated the distance of 1.3-1.6 kpc and the white dwarf mass of 0.58-0.63$M_{\odot}$. Poggiani~(2010) mentioned that the low white dwarf mass suggested that V5558 Sgr is a critical system whose mass is close to the lower limit to trigger the nova outburst.

Our spectral observations are summarized in section 2. The results of the spectral monitoring of V5558~Sgr are described in section 3. We discuss the properties in the light curves and the spectral evolutions in section 4. Conclusions are stated in section 5.

\section{Observation}
Our low dispersion spectra with a resolution of $\lambda / \Delta \lambda \approx 600$ (at 5852~\AA) have been taken with the FBSPEC1 and FBSPEC2, which have been developed by one of the authors (MF),  attached to the 28 cm telescope of the Fujii-Bisei Observatory (Okayama, Japan). The spectral coverage is 3800 to 8400 \AA. The spectra were obtained in 32 nights when the targets were brighter than V = 10 - 11 mag. The signal-to-noise ratio is typically 10 to 30. The reduction of the spectra was carried out with NOAO IRAF in the standard way, and the flux calibration was performed using a local spectrophotometric standard star.. Table \ref{tab_observation} gives the journal of our spectroscopic observations.

\begin{longtable}{llll}
  \caption{Spectroscopic Observations of Classical Nova V5558~Sgr}\label{tab_observation}
  \hline              
   Date & Specral range & JD & comments\\
         &  (\AA) & (2454000$+$) \\
\endfirsthead
  \hline
   Date & Specral range(\AA) & JD (2454000$+$) & comments\\ 
\endhead
  \hline
\endfoot
  \hline
\endlastfoot
  \hline
  2007 April 16 & 3800-8300 & 207.27 &\\
  \phantom{2007} April 25 & 3800-8300 & 216.27 &\\
  \phantom{2007} May 2 & 3800-8200 & 223.27 &\\
  \phantom{2007} May 8 & 3800-8200 & 229.27 &\\
  \phantom{2007} May 25 & 3800-8300 & 246.18 &\\
  \phantom{2007} June 10 & 3800-8200 & 262.16 &\\
  \phantom{2007} July 11 & 3800-8200 & 293.07 & at the initial maximum\\
  \phantom{2007} July 23 & 3800-8300 & 305.03 & after the initial maximum\\
  \phantom{2007} July 27 & 3800-8200 & 309.00 & around the second mini-maximum\\
  \phantom{2007} July 30 & 3800-8300 & 312.00 & around the second mini-maximum\\
  \phantom{2007} August 9 & 3800-8300 & 321.99 & around the second mini-maximum\\
  \phantom{2007} August 12 & 3800-8300 & 325.02 & at the third maximum\\
  \phantom{2007} August 15 & 3800-8200 & 327.98 &\\
  \phantom{2007} August 16 & 3800-8200 & 329.06 &\\
  \phantom{2007} August 25 & 3800-8300 & 337.97 &\\
  \phantom{2007} September 5 & 3800-8300 & 348.97 &\\
  \phantom{2007} September 12 & 3800-8200 & 355.95 &\\
  \phantom{2007} September 25 & 4600-6700 & 368.91 & at the fourth maximum\\
  \phantom{2007} October 6 & 4600-6700 & 379.91 &\\
  \phantom{2007} October 23 & 4600-6700 & 396.86 & at the fifth maximum\\
  \phantom{2007} November 2 & 4600-6700 & 406.92 & declining from the fifth maximum\\
  \phantom{2007} November 7 & 4600-6700 & 411.90 &\\
  \phantom{2007} November 11 & 4600-6700 & 415.89 &\\
  \phantom{2007} November 23 & 4600-6700 & 427.88 &\\
  2008 February 12 & 3900-8300 & 509.37 &\\
  \phantom{2008} February 27 & 3900-8400 & 524.34 &\\
  \phantom{2008} March 8 & 3900-8400 & 534.35 &\\
  \phantom{2008} March 17 & 3900-8400 & 543.32 &\\
  \phantom{2008} March 31 & 3900-8400 & 557.31 &\\
  \phantom{2008} April 14 & 3900-8400 & 571.28 &\\
  \phantom{2008} April 24 & 3900-8300 & 581.28 &\\
  \phantom{2008} May 3 & 3900-8300 & 590.26 &\\
\end{longtable}

\begin{figure}
  \begin{center}
    \FigureFile(80mm,80mm){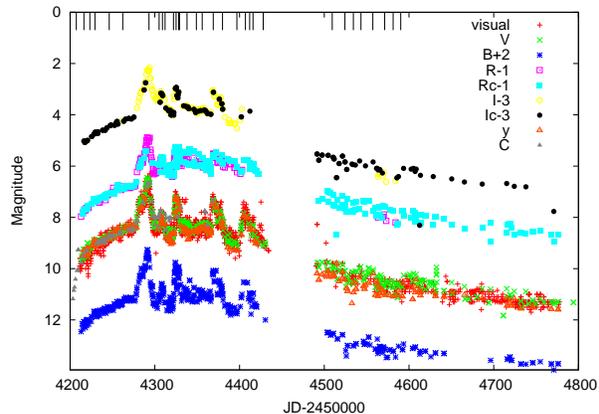}
  \end{center}
  \caption{Light curve of V5558~Sgr. The photometric data was taken from VSNET, AAVSO, ASAS-3 and Pi of the Sky. The ASAS-3 data
indicate that V5558 Sgr was fainter than than 12.2 until JD2454197. Although the contribution of a nearby star is included in the
original ASAS-3 data, we subtracted this contribution from the ASAS-3 data before drawing this figure. The epochs of our spectroscopic observations are marked with the tick marks.}\label{v55581}
\end{figure}

\section{Results}
Figure \ref{v55581} shows the light curve of V5558~Sgr collected from the archive of VSNET (\cite{Kato2004}), AAVSO\footnote{http://www.aavso.org/}, ASAS-3 (\cite{Pojmanski2002}) and Pi of the Sky archives\footnote{http://grb.fuw.edu.pl/pi/}. The brightness slowly rise for about two months before the initial brightness maximum. This flat phase is known as the pre-maximum halt. It is known that the duration of the pre-maximum halt depends on the decline speed of the nova (see \cite{Bode1989}). In slow novae, the duration is typically several days. The duration of the pre-maximum halt is, however, at least 60 day in V5558~Sgr, which is much longer than the typical duration. After the first maximum, the light curve shows several rebrightenings superposed on a flat part that lasts about four months until JD 2454405. The pre-maximum halt and repetitive outbursts were observed also in V723~Cas, V1548~Aql, HR~Del and DO~Aql (see e.g. V723~Cas: \cite{Ohsima1996, Munari1996, Iijima1998}; V1548~Aql: \cite{Kato2001}; HR~Del: \cite{Terzan1970}; DO~Aql: \cite{Vogt1928}). After the rebrightenings, V5558 Sgr decayed slowly and smoothly, though there are small fluctuations. In this section, we will present the spectra of V5558 Sgr in our observations during the three stages; the pre-maximum stage, the rebrightening stage, and the decline stage after the rebrightening stage.  

\subsection{The pre-maximum stage}
\begin{figure*}
 \begin{center}
   \FigureFile(160mm,50mm){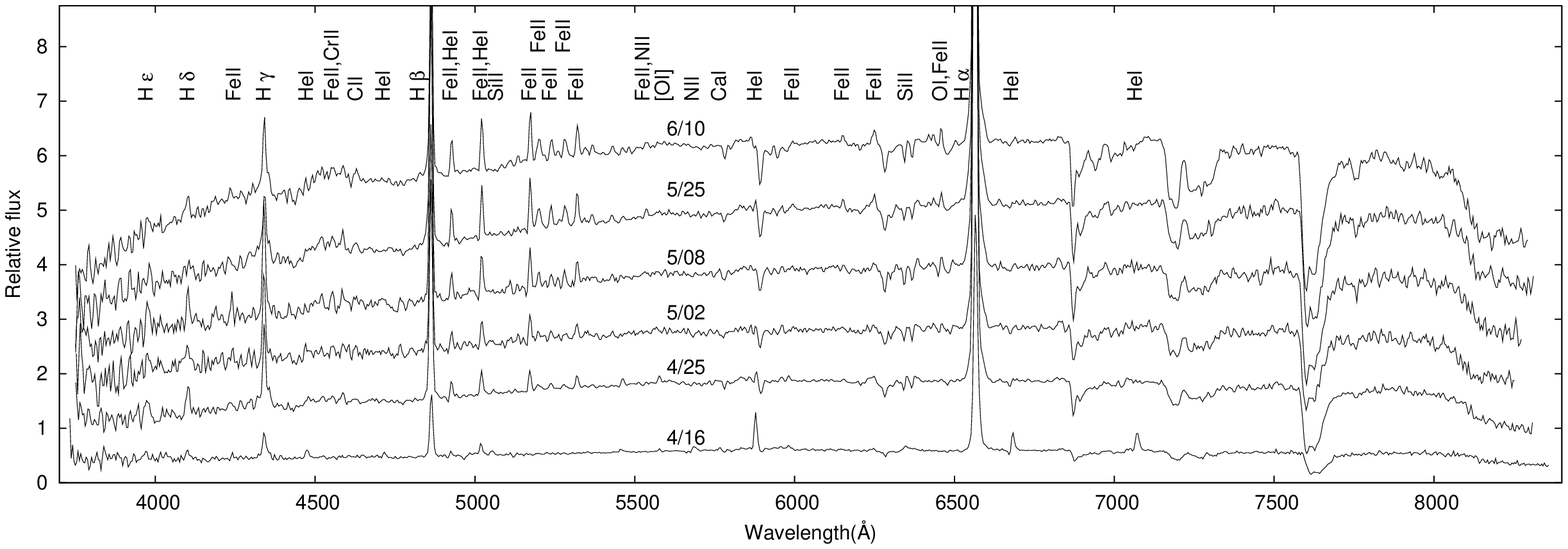}
  \end{center}
 \caption{Representative spectral evolution of V5558~Sgr during the pre-maximum phase from 2007 April 16 to June 10. For visuality, the data after April 25 are vertically shifted.}\label{v55582}
\end{figure*}

Figure \ref{v55582} shows the spectra of V5558~Sgr during the pre-maximum stage. In this stage, we obtained the spectra on 6 nights between 2007 April 16 and 2007 June 10 (JD 2454202.27 - 2454262.16). The first observation was carried out on 2007 April 16 (JD 2454202.27), two days after the discovery. This spectrum shows the presence of emission lines of Balmer, He~I, N~II, Si~II, Fe~II, Ca~I and [O~I], but is dominated by the strong Balmer and He~I lines. These emission lines exhibited sharp profiles, unlike the typical flat-topped profiles of He/N novae. FWHMs of these emission lines are about 500-700 km s$^{-1}$, which is almost equal to the typical value of Fe~II type novae, not He/N novae.

On April 25 (JD 2454216.27), He~I lines became weaker compared to the previous spectrum, while some Fe~II lines appeared in emission. Moreover, the emission line of SiII disappeared, and the absorption lines of Si~II~6347 and 6371 appeared instead. These Si~II absorption lines are also seen in the pre-maximum spectra of V723~Cas. In next 4 spectra between May 2 and June 10 (JD 2454223.27 - 2454262.16), there are strong Balmer and Fe~II lines. In addition, there are also weak C~I, C~II, O~I and [Fe~II] lines in emission, and the absorption lines of Si~II. The red part of these spectra shows absorption lines, due to atmosphere absorption.

\begin{figure}
 \begin{center}
   \FigureFile(80mm,50mm){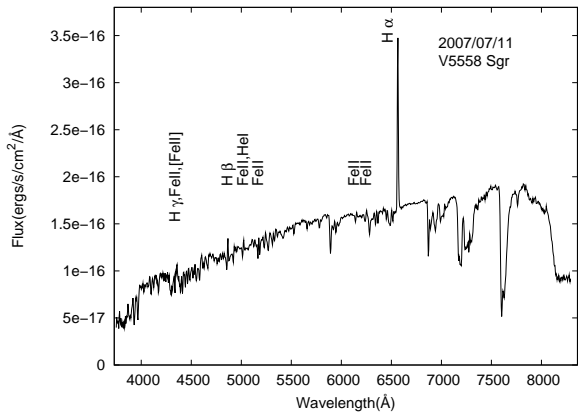}
  \end{center}
 \caption{Spectrum of V5558~Sgr on 2007 July 11.}\label{v55583}
\end{figure} 

\subsection{The rebrightenings stage}
\begin{figure*}
 \begin{center}
   \FigureFile(160mm,50mm){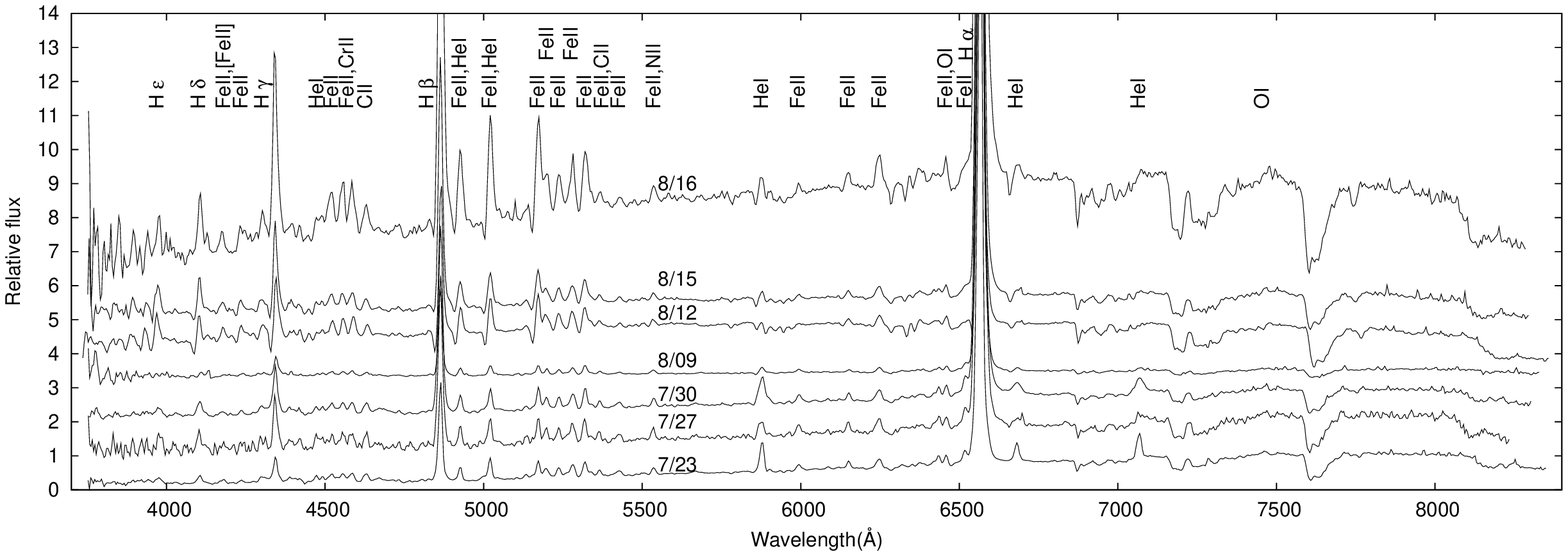}
  \end{center}
 \caption{Representative spectral evolution of V5558~Sgr during the early rebrightening stage from 2007 July 23 to August 16. For visuality, the data after July 27 are vertically shifted.}\label{v55584}
\end{figure*}

Next, we present the spectra during the rebrightening stage. On 2007 July 11 (JD 2454293.07), just at the initial maximum, the spectrum has remarkably changed (figure \ref{v55583}). There are Balmer and Fe~II lines in emission with an absorption component of the P-Cygni profile. All of the spectra between July 23 and August 16 are presented in figure \ref{v55584}. On July 23 (JD 2454305.03), after the initial maximum brightness, there are many emission lines of Balmer, Fe~II, He~I, O~I, N~II, C~II, Cr~II and [Fe~II]. In the next three observations (JD 2454309.00 - 2454321.99), at the second mini outburst and before the third maximum brightness, presented in figure \ref{v55584}, the spectra do not show distinct change compared to the previous spectrum regarding identified lines.

The next spectrum was obtained on August 12 (JD 2454325.02), at the third maximum. We can identify the Balmer series, Fe~II, O~I, [O~I], N~II, C~II and He~I lines. In Balmer series and Fe~II lines, we can identify strong P-Cygni profiles. Moreover, the continuum is stronger than the previous spectrum. The spectra on August 15 and 16 (JD 2454327.98 and 2454329.06) are very similar to the previous spectrum regarding identified lines, though the absorption component of the P-Cygni profile becomes weak and He~I lines become strong.

\begin{figure*}
 \begin{center}
   \FigureFile(160mm,50mm){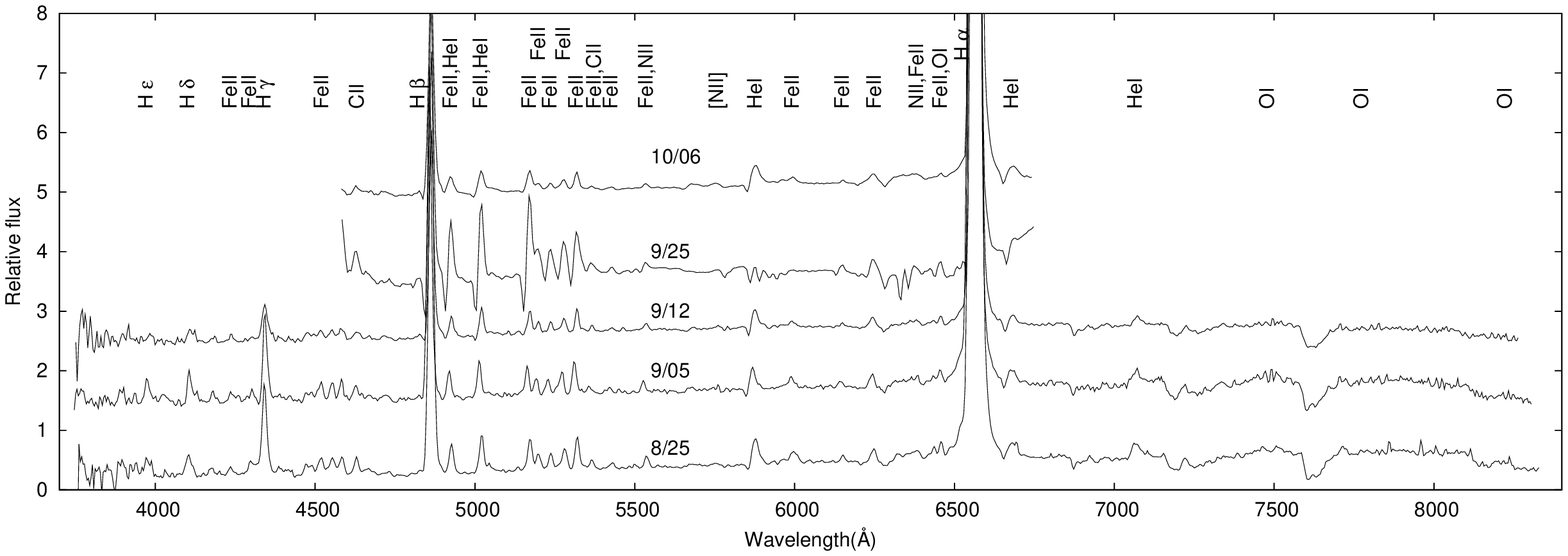}
  \end{center}
 \caption{Representative spectral evolution of V5558~Sgr during the mid rebrightening stage from 2007 August 25 to October 6. For visuality, the data after September 5 are vertically shifted.}\label{v55585}
\end{figure*}

During the plateau phase between the third and fourth brightness maximum, we observed this object on 3 nights (JD 2454337.97 - 2454355.95) (figure \ref{v55585}). There are Balmer, Fe~II, O~I, He~I, N~II, C~II and [Fe~II] lines in emission on all of the 3 nights, and the absorption component of the P-Cygni profile appears in He~I~5876. In addition, on September 12 (JD 2454355.95), we can identify [N~II]~5755.

On September 25 (JD 2454368.91), just at the fourth maximum, the spectrum had obviously changed; the continuum became strong, and the P-Cygni profile reappeared in Balmer and Fe~II lines. After the peak, V5558 Sgr rapidly decayed. On October 6 and 23 (JD 2454379.91 and 2454396.86), we can also identify P-Cygni profiles in Balmer and Fe~II lines. He~I~5876 becomes strong and is accompanied by an absorption component of the P-Cygni profile. There are [N~II] and [O~I] lines, too.

\begin{figure*}
 \begin{center}
   \FigureFile(160mm,50mm){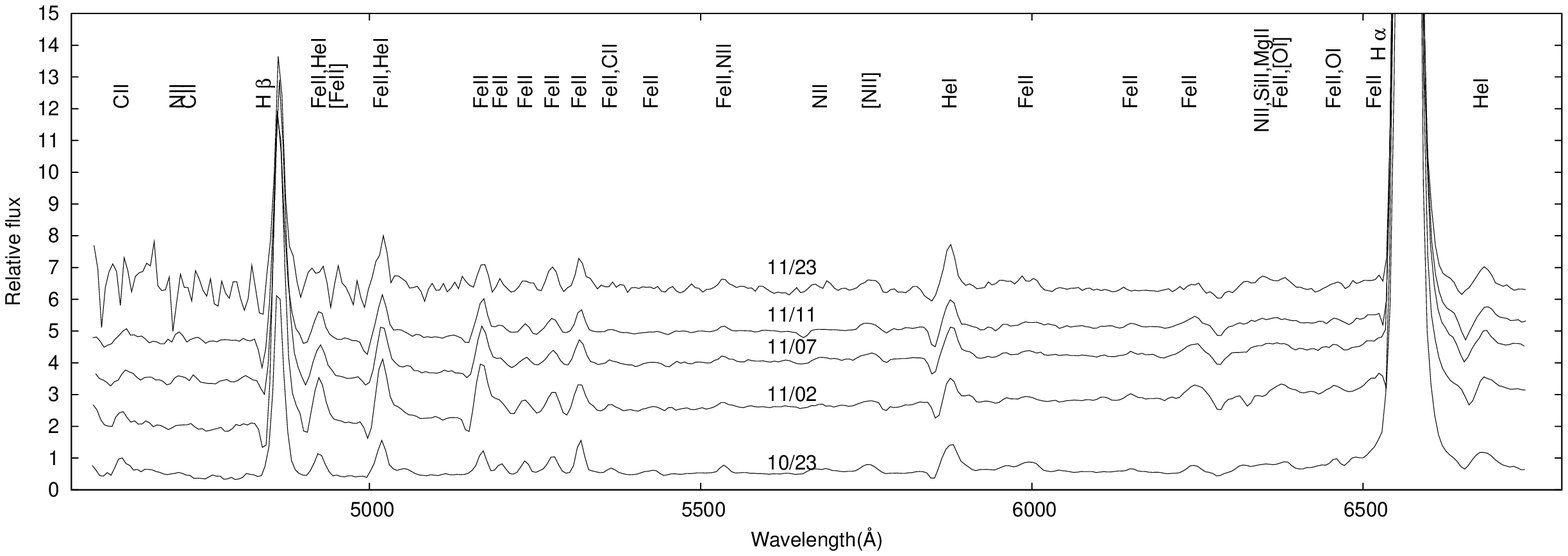}
  \end{center}
 \caption{Representative spectral evolution of V5558~Sgr during the late rebrightening stage from 2007 October 23 to November 23. For visuality, the data after November 2 are vertically shifted.}\label{v55586}
\end{figure*}

The spectra during the later phase of the rebrightening stage were observed at the fifth maximum and later (figure \ref{v55586}). On November 2 (JD 2454406.92), There are Balmer series, Fe~II, He~I, O~I, [O~I], N~II, [N~II] and C~II lines in emission. P-Cygni profiles appear in Balmer and Fe~II lines, in addition to He~I line. In the next 3 spectra, on November 7, 11 and 23 (JD 2454411.90 - 2454427.88), we identify the similar characteristics in emission. The absorption components of P-Cygni profiles decayed by little and little.

\subsection{The decline stage after the rebrightening stage}
\begin{figure*}
 \begin{center}
   \FigureFile(160mm,50mm){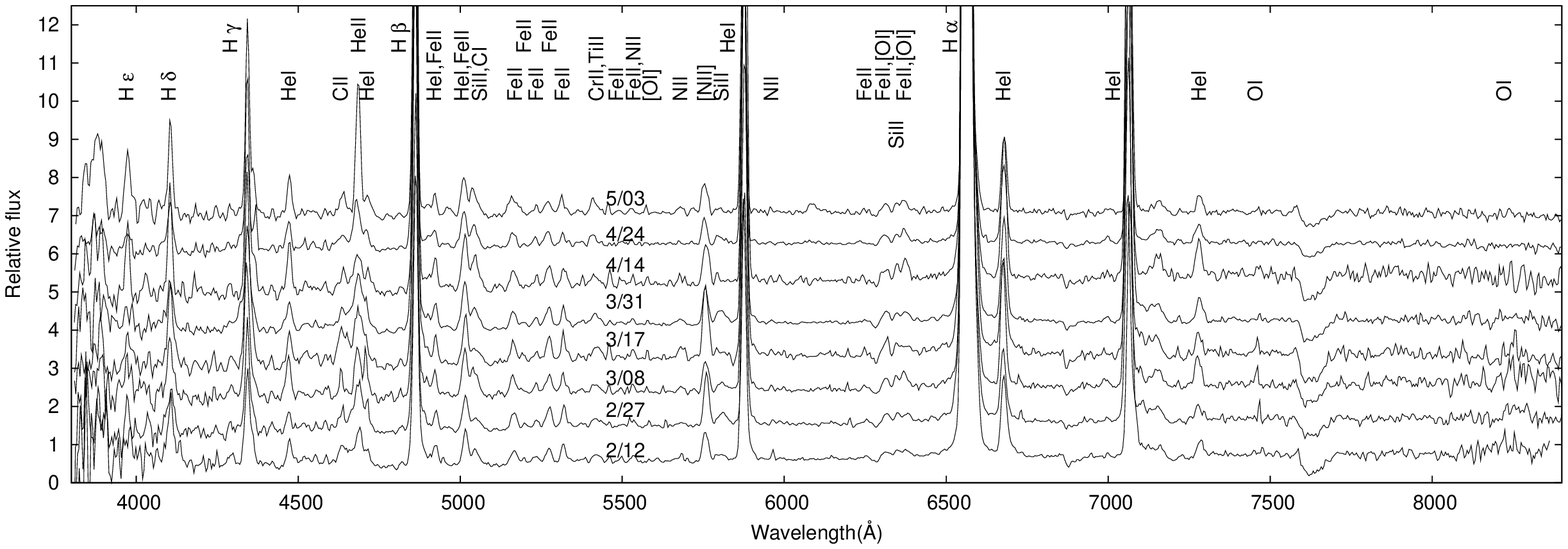}
  \end{center}
 \caption{Representative spectral evolution of V5558~Sgr during the decline stage from 2008 February 12 to May 3. For visuality, the data after February 27 are vertically shifted.}\label{v55587}
\end{figure*}

After the fifth maximum, the light curve declined smoothly. V5558 Sgr were, however, unobservable between 2007 November 28 and 2008 January 29, due to the solar conjunction. In this subsection, we present the spectra obtained after this non-observation phase. All of the spectra during the decline stage are presented in figure \ref{v55587}. In this stage, the intensities of He~I lines increased, compared to the previous stage. On February 12 (JD 2454509.37), there are Balmer, He~I, N~II, Fe~II, He~II, [N~II], [O~I], O~I, C~II and Si~II lines in emission. We can classify this nova as a hybrid nova, which evolved from the Fe~II nova toward the He/N nova. Note that He~I lines were dominant on 2007 April 16 (JD 2454202.27). On February 27 (JD 2454524.34), the main features are same as on the previous day, but the spectrum shows O~I~7454 and Si~II~5800 lines in emission, and disappearance of N~II~5680 and N~II~5961 lines. On March 8 (JD 2454534.35), there are Si~II or C~I lines (5041\AA) in emission. On May 3 (JD 2454590.26), the intensity of He~II~4686 increased, compared to the previous spectrum. There is  Cr~II or additional Fe~II lines (6082\AA or 6084\AA, respectively) in emission. 

\begin{longtable}{llllllllllllll}
  \caption{Blue-shift velocities in $\mathrm{km\;s}^{-1}$ of absorption components of the P-Cygni profiles. The error is typically $\pm$100 km s$^{-1}$. AI and RW are an atomic identification and a rest wavelength, respectively.}\label{tab_pcygni}
 \hline              
\endfirsthead
 \hline
\endhead
  \hline
\endfoot
  \hline
\endlastfoot
  \hline
  AI & RW (\AA) & 4/16 & 4/25 & 5/8 & 7/11 & 8/9 & 8/12 & 8/15 & 8/16 & 8/25 & 9/5 & 9/12\\
  \hline
  H$\gamma$ & 4340 & & & & & & $-1230$ & $-1260$ & & & $-1410$ &\\
  H$\beta$ & 4861 & & & & $-670$ & & $-1170$ & $-1370$ & $-1450$ & & $-1420$ & $-1410$\\
  Fe~II & 4924 & & $-620$ & & $-640$ & & $-1050$ & $-1150$ & $-1200$  & & &\\
  Fe~II & 5018 & & & $-680$ & $-640$ & & $-1050$ & $-1230$ & $-1160$ & & & $-1190$\\
  Fe~II & 5169 & & & $-680$ & & & $-1060$ & $-1240$ & $-1210$ & & &\\
  He~I & 5876 & & & & & $-1020$ & & $-970$ & $-920$ & $-1220$ & $-1060$ & $-1050$\\
  H$\alpha$ & 6563 & & & & & & $-1270$ & & &\\
  He~I & 6678 & $-540$ & & & & & & & $-1370$ & $-1120$ & $-1110$\\
  \hline
  \hline
  AI & RW (\AA) & 9/25 & 10/6 & 10/23 & 11/2 & 11/7 & 11/11 & 11/23 & 3/17\\
  \hline
  H$\gamma$ & 4340 & & & & & & & & \\
  H$\beta$ & 4861 & $-1240$ & $-1650$ & & $-1350$ & $-1530$ & $-1550$ & $-1700$ & \\
  Fe~II & 4924 & $-1090$ & & & $-1270$ & & & & \\
  Fe~II & 5018 & $-1030$ & $-1560$ & $-1360$ & $-1300$ & $-1460$ & $-1500$ & & \\
  Fe~II & 5169 & $-1110$ & $-1370$ & & $-1290$ & $-1390$ & $-1340$ & & \\
  He~I & 5876 & & $-1430$ & $-1410$ & $-1090$ & $-1420$ & $-1330$ & $-1500$ & $-1760$\\
  H$\alpha$ & 6563 & $-1330$ & & & $-1580$ & & $-1720$ & & \\
  He~I & 6678 & & $-1360$ & & $-1070$ & $-1400$ & $-1470$ & $-1520$ & \\
  \hline
\end{longtable}

\section{Discussion}
\subsection{The behavior during the pre-maximum stage and comparison with pre-maximum halts of other novae}
The light curve of V5558~Sgr shows a prominent premaximum halt, lasting for at least 60 days. Although the origin of the pre-maximum halt is not yet clear, it is known that the duration depends on the speed class. This long duration is also seen in some slow novae, V723~Cas, V1548~Aql, HR~Del and V463~Sct, known to be the very slow novae except V463~Sct. The evolution of the light curve of V5558~Sgr is similar to that of, especially, V723~Cas in that they both show the long pre-maximum halt, several rebrightenings on a flat part and a slow decline. V5558~Sgr is also classified as the very slow nova. This is compatible with the feature of the long pre-maximum halt and subsequent flare-like maxima.

Friedjung (1992) referred to the spectroscopic observations of HR~Del during the pre-maximum stage (cf. \cite{Hutchings1970}). They mentioned that the expansion velocity of the photosphere is vely low (200 km s$^{-1}$), unlike the majority of classical novae. They suggested that the conditions of thermonuclear runaway were only marginally satisfied in HR~Del. This interpretation is consistent with the derived white dwarf mass in HR~Del (0.52 $M_{\odot}$: \cite{Bruch1982}, 0.595 $M_{\odot}$: \cite{Kuerster1988}) according to the models of thermonuclear runaways (see \cite{Prialnik1995}). This suggestion may, thus, apply to the similar slow nova, V5558~Sgr. It is important to measure the white dwarf mass of V5558 Sgr. Note that Orio and Shaviv (1993) suggested another hypothesis, the local thermonuclear runaway, for the pre-maximum halt.

Kato et al.~(2002) advocated that V463~Sct provided a unique opportunity for testing the interpretation of pre-maximum halts, which requires that the conditions of thermonuclear runaway were only marginally satisfied.
Hachisu and Kato~(2004) presented an interpretation of long maximum halts from the theoretical side. They estimated the WD masses to be 0.59 $M_{\odot}$ for V723~Cas and 1.1 $M_{\odot}$ for V463~Sct by fitting the decline rate of reproduced light curves, excluding the brightness maxima, based on the optically thick wind model. They concluded that a nova attains its maximum brightness in the optical light and has a flat peak when the envelope mass is massive enough and the temperature of white dwarf surface decreases to below $\sim$7000 K. The duration of the flat peak is considered to depend on the initial envelope mass at the ignition (\cite{Hachisu2004}). This flat peak can be regarded as a premaximum halt. They interpreted several strong peaks of the light curve of V723 Cas as pulsations of the nova envelope as discussed by Schenker~(1999). Schenker~(1999) attempted to apply radial pulsations, based on the $\kappa$-mechanism, in envelopes to the physical origin of the oscillation phase of classical novae. In V5558 Sgr, the brightness during the pre-maximum halt slowly rise unlike V723 Cas. This feature is not explained by the scenario proposed by Hachisu and Kato (2004).

Iijima et al. (1998) presented the spectral evolution of V723~Cas. They reported the spectra in the long pre-maximum stage. There were Balmer and Fe~II lines in emission. Most of the emission lines were accompanied by strong absorption components of the P-Cygni profiles. The emission lines gradually weakened, and the absorption lines developed. These features of emission and absorption lines were not seen in our observations of V5558~Sgr. In our observations, there were also Balmer and Fe~II lines in emission. The emission lines, however, did not weaken, and strong absorption components of the P-Cygni profiles did not appear during the pre-maximum stage. On April 16 (JD 2454202.27), we can identify the absorption component of the P-Cygni profile of He~I~6678 with the blue-shift velocity, about $-500$ km s$^{-1}$. This value is higher than the velocities in V723~Cas, by about $-100$ km s$^{-1}$ or in HR~Del, by about $-200$ km s$^{-1}$ (see \cite{Iijima1998}, \cite{Friedjung1992}). Moreover, Iijima et al. (1998) reported that the blue-shift of the absorption components decreased with time during the pre-maximum stage and then slightly increased at the maximum brighteness. They suggested that this is because a new high velocity absorption system emerged. We can not confirm this trend in V5558~Sgr, since absorption components of the P-Cygni profiles disappeared during the pre-maximum stage. The detailed description of the spectral evolution is given in the subsection 4.3. The difference of V5558 Sgr and V723 Cas and HR Del suggests a possibility that V5558 Sgr has the heavier white dwarf and the ejecta of V5558 Sgr was lighter, and the specific photosphere could not form during the pre-maximum phase, then.

\subsection{Blue-shift velocities of the absorption components of the P-Cygni profiles}
Table \ref{tab_pcygni} shows the blue-shift velocities of the absorption components of the P-Cygni profiles. We hereafter call the velocity of the absorption component measured from the emission peak as the radial velocity for simplicity. In the pre-maximum stage, the radial velocities of He~I, Fe~II and H~I lines is about $-500$ to $-700$ km s$^{-1}$. After the initial brightness maximum, the radial velocities increase. In the rebrightening stage, the radial velocities become higher, over about $-1000$ km s$^{-1}$. we can find the trend that the radial velocity becomes low at the brightness maximum, and becomes high outside the brightness maximum. After the rebrightening stage, the absorption components disappear and we can not derive the radial velocity except on 2008 March 17. The radial velocity of He~I~5876 on March 17 is higher, $-1760$ km s$^{-1}$. The long-term trend of increase of the radial velocity suggests that the velocity is higher in the inner region because the photosphere gradually shrinks. This suggestion is consistent with the low radial velocity at the brightness maximum when re-expansion of the photosphere ocuurs as noted in the subsection 4.4.

\subsection{Spectral type of V5558 Sgr}
The spectral evolution of V5558 Sgr is very rare. On 2007 April 16 (JD 2454202.27), during the pre-maximum halt, the spectrum of V5558 Sgr is dominated by Balmer and He~I lines in emission. These emission lines of the spectrum show the nature of the He/N type nova. The FMHWs of H$\alpha$ and He~I~5876 \AA \ are, however, of low velocities, about 700 km s$^{-1}$ and 500 km s$^{-1}$, respectively, unlike in the majority of He/N novae. He~I lines then weaken very much by April 25 and Fe~II lines appear. Williams (1992) reported existense of the hybrid type nova. They suggested that Fe~II spectra, in the transition from the Fe~II to the He/N spectrum, with unusually broad lines, the half width at zero intensity (HWZI) $>$ 2500 km s$^{-1}$, probably originate from discrete shells which are rather massive and therefore optically thick. As the expanding shell becomes optically thin, the spectrum changes to the more typical higher ionization He/N spectrum of the shells. This scenario may not apply the pre-maximum stage. The spectrum during the pre-maximum halt of slow novae is basically dominated by Fe~II type emission lines (see figure \ref{v55582} in this paper and \cite{Iijima1998}). The spectrum at the earliest phase during the pre-maximum halt may also show emission lines of the He/N type also in other novae. The earliest spectrum may have been overlooked in other novae since the duration of this stage should be very short. The spectral transition of V5558 Sgr during the early stage, from the He/N type toward the Fe~II type, is a rare evolution at least at this time. 
Moreover, the spectra of V5558 Sgr then evolved to the Fe~II and He/N hybrid type after the initial maximum brightness (for example, the spectrum on 2007 July 23). We note that He~I lines become weak and P-Cygni absorptions become strong at the maximum brightness. Futhermore, He~I and He~II lines become strong at the decline stage after the rebrightening stage. FWHMs of He~I or Balmer at the decline stage are about 800-900 km s$^{-1}$, which is slower than the typical He/N novae.

As noted in the subsection 4.1, Iijima et al. (1998) presented the spectral evolution of V723~Cas. They reported the spectra in the long pre-maximum stage. There were Balmer and Fe~II lines in emission. Most of the emission lines were accompanied by strong absorption components of the P-Cygni profiles.
Iijima (2006) reported on the spectral evolution after the maximum brightness of V723 Cas. In V723 Cas, two weeks after its maximum and at about two magnitudes below its maximum, there are Balmer, Fe~II and He~I emission lines, simultaneously. At the second brightening, P-Cygni absorption components appeared on Balmer, He~I, and Fe~II lines. When the nova faded by about four magnitudes, during the decline stage, the prominent emission lines are of Balmer, He~I, [N~II], [Fe~II], [O~III], N~II, N~III, and Si~II, while most of the emission lines of Fe II have disappeared. This transition after the initial maximum brightness of V723 Cas is similar to that of V5558 Sgr. The transition during the pre-maximum stage are, however, different between V5558 Sgr, He/N type toward Fe~II type, and V723 Cas. It should be noted that V723 Cas possibly showed the He/N spectra before the first observation by Iijima et al. (1998).

\subsection{Comparison with the rebrightening novae during the early phase}
The light curves of some novae, for example V1186~Sco, V2540~Oph, V5113~Sgr, V4745~Sgr and V458~Vul, show several rebrightenings during the early phase (see e.g. V1186~Sco: \cite{Schwarz2007}; V2540~Oph: \cite{Ak2005}; V5113~Sgr: \cite{Kiyota2004}; V4745~Sgr: \cite{Csak2005}; V458~Vul: \cite{Poggiani2008b}). The spectra of these novae show many emission lines accompanied with the absorption components of P-Cygni profiles at the brightness maximum (\cite{Tanaka2011}). Tanaka et al.~(2011) suggested that the re-appearance of the absorption component at the rebrightening maxima is attributable to re-expansion of the photosphere after it once shifts sufficiently inside. The spectra of V5558~Sgr also show the similar trend, that the absorption components of P-Cygni profiles develop at the brightness maximum. The photosphere of V5558~Sgr could, thus, cause the similar physical transition. They also reported that the time interval between successive maxima increase gradually. We can not identify this trend for V5558 Sgr.

\section{Conclusions}
The light curve of V5558~Sgr shows the pre-maximum halt and then several flare-like rebrightenings. V5558~Sgr is classified as a very slow nova, and is similar to V723~Cas. We obtained the low-resolution spectra of V5558 ~Sgr at the Fujii-Bisei Observatory on 32 nights between 2007 April 16 and 2008 May 3.

We presented the behavior of the pre-maximum stage and compared with other novae. The transition of spectra of V5558 Sgr is similar to the very slow nova V723 Cas roughly. The transition from He/N type to Fe~II type during the pre-maximum stage, however, seems a unique phenomenon in V5558 Sgr, though there is a possibility that this transition is undergone also in other novae, but have been missed since it occurs well before the maximum.

We found the trend that the radial velocity of the blueshift absorption component from the emission peak gradually increases, but becomes low at the brightness maximum.

Our data revealed spectroscopic changes during the pre-maximum halt and then several flare-like rebrightenings. The spectra at the pre-maximum halt shows the rare evolution. There are Balmer and He~I lines at the first observation (JD 2454202.27), and then the spectra varied from the He/N nova to the Fe~II type nova. The spectra at and around brightness maximum during the rebrightenings, we can identified many lines accompanied by a strong absorption component of the P-Cygni profile. There are also He~I and Fe~II lines simultaneously. Moreover, He~I and He~II lines become strong during the decline stage. FWHMs of emission lines are, however, lower than that of typical He/N novae.

In addition, we compared V5558 Sgr with other novae showing the several rebrightenings during the early phase. There are common features that the absorption components of the P-Cygni profiles become strong at the maximum brightness. There is a possibility that the photosphere reexpand at maximum brightness.

\bigskip
The authors are thankful for observers who have posted their precious data to VSNET and AAVSO. We used the data which were obtained by the ASAS-3 system and Pi of the sky, and made publicly available. This work was supported by the Grant-in-Aid for the Global COE Program "The Next Generation of Physics, Spun from Universality and Emergence" from the Ministry of Education, Culture, Sports, Science and Technology (MEXT) of Japan.

\end{document}